\documentclass[aps,prl,twocolumn,superscriptaddress,showpacs]{revtex4}
\usepackage{feyn}
\usepackage{graphicx}
\usepackage{latexsym}
\usepackage{amssymb}
\usepackage{amsmath}
\usepackage{amsfonts}
\usepackage{bm}
\usepackage{multirow}
\usepackage{color}
\usepackage{comment}

\newcommand{\cB}{\mathcal{B}}
\newcommand{\cA}{\mathcal{A}}

\newcommand{\hn}{\hat{n}}

\newcommand{\bra}[1]{\langle #1|}
\newcommand{\ket}[1]{|#1 \rangle}

\newcommand{\beq}{\begin{equation}}
\newcommand{\eeq}{\end{equation}}
\newcommand{\beqn}{\begin{eqnarray}}
\newcommand{\eeqn}{\end{eqnarray}}

\DeclareMathAlphabet{\mathbbold}{U}{bbold}{m}{n}

\newcommand{\mcal}{\mathcal}

\usepackage{ulem}

\begin{document}

\title{Construction of Fractal Order and Phase Transition with Rydberg Atoms}

\author{Nayan E. Myerson-Jain}

\author{Stephen Yan}

\author{David Weld}

\author{Cenke Xu}

\affiliation{Department of Physics, University of California,
Santa Barbara, CA 93106, USA}

\begin{abstract}

We propose the construction of a many-body phase of matter with
fractal structure using arrays of Rydberg atoms. The degenerate
low energy excited states of this phase form a self-similar
fractal structure. This phase is analogous to the so-called
``type-II fracton topological states". The main challenge in
realizing fracton-like models in standard condensed matter
platforms is the creation of multi-spin interactions, since
realistic systems are typically dominated by two-body
interactions. In this work, we demonstrate that the Van der Waals
interaction and experimental tunability of Rydberg-based platforms
enable the simulation of exotic phases of matter with fractal
structures, and the study of a quantum phase transition involving
a fractal ordered phase.

\end{abstract}

\maketitle

%{\it --- Introduction}

In recent years, tremendous progress has been made in simulating
quantum many-body systems with highly tunable arrays of Rydberg
atoms~\cite{rydbergreviewbrowaeys,gross2017quantum,saffman2010quantum}.
In many such experiments, the ground state and a high-lying
excited state of the atom constitute a qubit, the fundamental
element of numerous exotic quantum many-body
states~\cite{wenzoo,subirreview}. Recently, the construction of
unconventional many-body states like $Z_2$ quantum spin liquids
has been explored~\cite{Rydbergashvinz2,Rydbergsubirz2,topoex},
demonstrating the highly promising potential of the Rydberg-based
platforms. The possibility to extend these platforms to realize
quantum many-body systems beyond the currently well-understood
theoretical paradigm such as those exhibiting $Z_2$ topological
order~\cite{subirz2,wenz2,kitaevz2} would be extremely exciting.

``Fracton" phases of matter provide a natural playground for
exotic physics. These phases host excitations with restricted
dynamics, and a ground-state degeneracy that scales with the
system
size~\cite{Haahcode,ChamonTypeI,BravyiTypeI,yoshida,Sagarclassification,Sagarduality,fractonreview1,fractonreview2}.
Fracton related models are loosely classified by their qualitative
features: ``type-I" models have excitations whose dynamics are
restricted to standard submanifolds, e.g lines and planes in
space, while excitations of the more exotic ``type-II" models are
created at the end of a fractal subset of the lattice and are
completely immobile~\cite{Sagarduality}. The prototypical example
of the type-II model is Haah's code~\cite{Haahcode}, which is
defined on the cubic lattice and has a ground-state degeneracy
which scales subextensively with the system size. While fracton
related phases are of great theoretical interest, as they
represent a world beyond well-studied topological quantum field
theories, much less progress has been made in realizing these
models experimentally.

In this work, we propose an experimental realization of a
two-dimensional analogue of a type-II fracton phase, as well as a
quantum phase transition between a phase with ``fractal order" and
a trivial phase. The fractal order spontaneously breaks a fractal
subsystem symmetry, and its low energy excitations form a
Sierpinski triangle on the lattice, which is a fractal shape with
Hausdorff dimension $d_H = \ln 3/\ln 2$, and only costs energy at
the corners of the Sierpinski triangle. The corners of the
triangle are also completely immobile, as their motion inevitably
creates higher energy excitations. We stress that the fractal
order we consider is defined as spontaneous breaking of a fractal
subsystem symmetry; this phase does not have topological order as
in Haah's code~\cite{Haahcode}.

{\it --- Sierpinski Triangle Model}

\begin{figure}
\begin{center}
\includegraphics[width=0.45\textwidth]{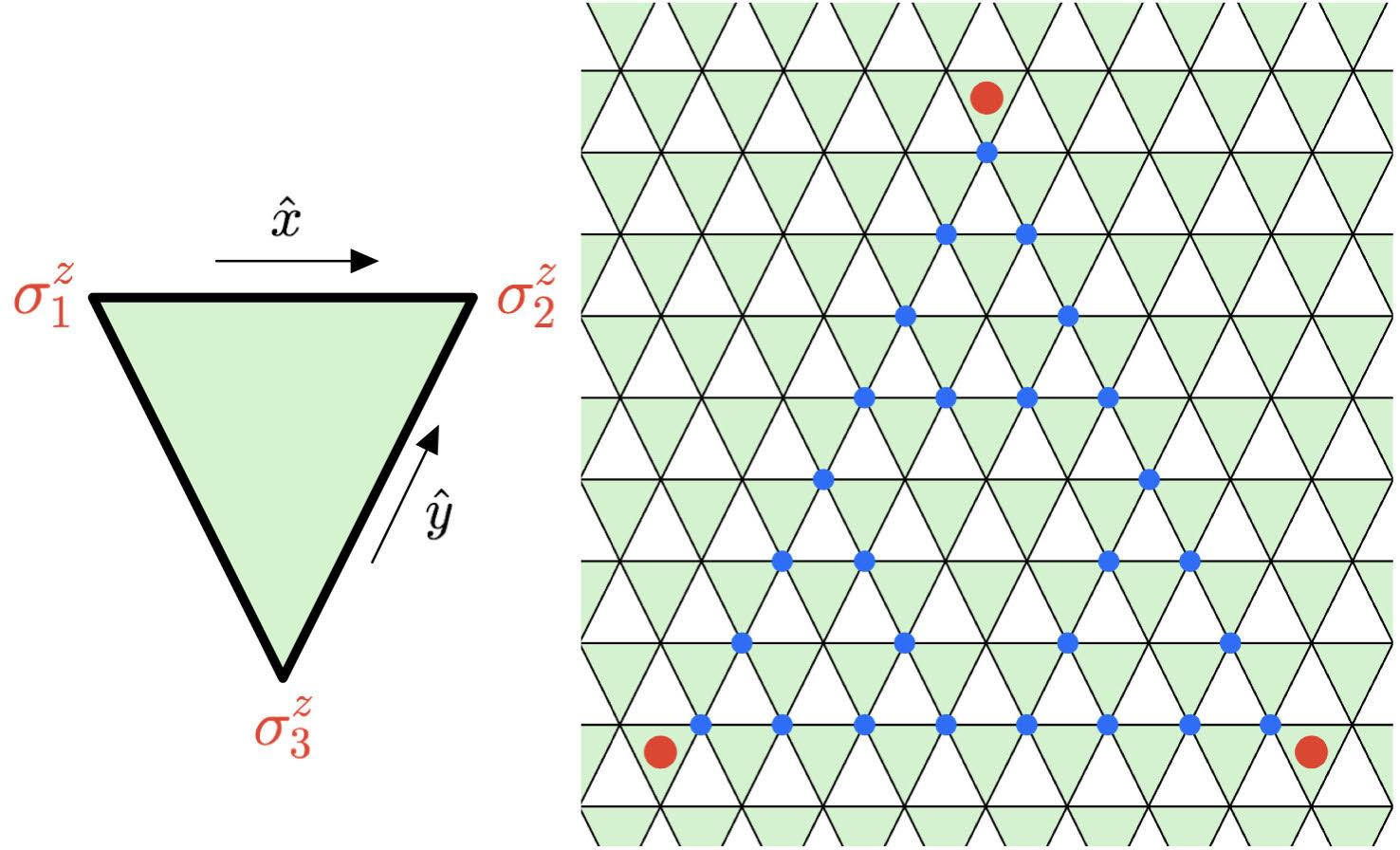}
\caption{(Left): Three-body spin interaction on each downward
facing triangle in Eq.~\ref{st}. (Right): one of the low energy
excitations of Eq.~\ref{st}; starting with the obvious ground
state with $\sigma^z = +1$, the spins are flipped to $\sigma^z =
-1$ on a $\ell = 3$ Sierpinski triangle (labelled blue), this
configuration only costs energy on the unit triangles at the three
corners (labelled red).}\label{ST_model}
\end{center}
\end{figure}

The Sierpinski triangle model~\cite{NM,yoshida} is the
paradigmatic model with fractal order. It is a classical
statistical mechanical model for an Ising system on a triangular
lattice whose Hamiltonian takes a simple form: one sums over all
three-body interactions on downward facing unit triangles
$\bigtriangledown$ as opposed to upward facing triangles
$\triangle$ (Fig.~\ref{ST_model}):
\begin{equation}
H_{\mathrm{ST}} = \sum_{\bigtriangledown} - K \sigma^z_1
\sigma^z_2 \sigma^z_3, \label{st}
\end{equation}
where the $\sigma^z = \pm 1$ are the Ising degrees of freedom
placed at the vertices of each downward unit triangle. The low
energy excited states of this model have a fractal structure:
starting with the obvious ground state with uniform $\sigma^z_j =
+1$, the low energy excited states correspond to flipping spins in
the shape of a Sierpinski triangle, which does not cost energy
anywhere except for the three corners of the Sierpinski triangle.
If we view the corners of the Sierpinski triangle as point
particles with finite energy, these particles cannot move along
the lattice without creating more excitations that cost higher
energy. Hence the mobility of these particles is highly
restricted, and it is in this sense that they are fractons.

A quantum version of the Sierpinski triangle model was discussed
in Ref.~\cite{juan1,frankyizhi}. Just like the classic quantum
Ising model, the quantum Sierpinski triangle model has an extra
transverse field \beqn H_{q\mathrm{ST}} = \sum_{\bigtriangledown}
- K \sigma^z_1 \sigma^z_2 \sigma^z_3 - \sum_j h \sigma^x_j.
\label{qst} \eeqn It was shown that the quantum Sierpinski
triangle model has two highly desirable features:

{\it (1)} It is ``self-dual", meaning that under a duality
transformation the $K$ and $h$ term will exchange. This
self-duality is analogous to that of the Kramers-Wannier duality
of the $1d$ quantum Ising model~\cite{wannier,kogut}, and the
self-duality of the quantum plaquette model on the square
lattice~\cite{xumoore}. The self-duality implies that, if there
were a quantum phase transition reached by tuning $h/K$ in
Eq.~\ref{qst}, it must happen at $h = K$.

{\it (2)} Numerical simulation of the quantum Sierpinski triangle
model suggests that the system may have a second order quantum
phase transition at the self-dual point $h = K$~\cite{frankyizhi}
(though earlier numerics suggested a first order
transition~\cite{juan1}); at the quantum phase transition, the
energy density has a fractal dimension $d_H = \ln 3 / \ln 2$
rather than scaling dimension 2 as in ordinary quantum phase
transitions in $2d$~\cite{frankyizhi}. This quantum phase
transition is associated with the spontaneous breaking of a
``fractal symmetry"; the phase with $h < K$ is identified as a
``fractal order", while the phase with $h > K$ is a disordered
phase of the fractal symmetry (see the appendix for more
discussion).

The nature of the quantum phase transition at $h = K$ in
Eq.~\ref{qst} is far from being understood, and the ordinary
Landau-Ginzburg paradigm is no longer applicable. Numerics suggest
that this transition is likely continuous, but many questions
await answers. For example: is the continuous quantum phase
transition stable against perturbations? For ordinary quantum
phase transitions, this question is answered through the
well-established renormalization group (RG)
method~\cite{wilson1,wilson2,wilson3}, by evaluating the relevance
or irrelevance of certain perturbations. But for the quantum phase
transition under discussion, no reliable RG procedure has been
established. Hence, key aspects of the transition must be explored
experimentally. A tunable experimental realization of the
classical and quantum Sierpinski triangle model would be extremely
useful in understanding quantum phase transitions involving
fractal geometry.

{\it --- Realizing the Sierpinski Triangle model with Rydberg
atoms}

The goal of this work is to describe a construction of both the
classical and quantum Sierpinski triangle models from arrays of
Rydberg atoms. To begin, let us consider a single atom for which
we couple the ground state $\ket{g}$ and an excited Rydberg state
$\ket{r}$ via a laser detuned from resonance. The two states
coupled by the laser are the atom-field product states labelled
$\ket{g, N_\gamma+1}$ and $\ket{r, N_\gamma}$, where $N_\gamma$ is
the photon number of the laser so that $\ket{g,N_\gamma+1}$ has
one extra photon compared to $\ket{r, N_\gamma}$. In the effective
two-state problem, the Rabi frequency enters as a term coupling
these two states. The simplest manifestation of the Rabi
oscillations is as a term in the Hamiltonian $\Omega \sigma^x$
where $\sigma^x = \ket{g, N_\gamma+1}\bra{r, N_\gamma} + \ket{r,
N_\gamma}\bra{g, N_\gamma+1}$.

If we blue-detune the laser from resonance, the energy gained by
the atom being in the excited state $\ket{r, N_\gamma}$ relative
to being in the state $\ket{g, N_\gamma+1}$ is $-\delta$, where
$\delta$ is the detuning of the laser. This detuning then
contributes a diagonal term to the effective Hamiltonian $- \delta
\hn$ where $\hn$ is $0$ if the atom is in the state $\ket{g,
N_\gamma+1}$ and $1$ if the atom is in the state $\ket{r,
N_\gamma}$. This allows us to write down an effective two-state
Hamiltonian in the basis of atom-field product states for the
single atom
\begin{equation}
H_{1 \text{ atom}} = \Omega \sigma^x - \delta \hn. \label{rabi}
\end{equation}

Two atoms in $s-$orbital Rydberg states interact through a force
that can be modelled by a Van der Waals potential $V(r) = C/r^6$
when the inter-atomic separation $r$ is large, where $C$ is a
constant that scales very strongly with the principal quantum
numbers of the two Rydberg states. For two identical Rydberg atoms
with principal quantum number $n$ (not to be confused with the
number operator $\hn$), the coefficient $C$ of the Van der Waals
interaction roughly scales with $\sim n^{11}$. In the remaining of
the paper we will use the more detailed evaluation of the Van der
Waals interaction given in Ref.~\cite{Sibalic_2017}. As such, the
total effective many-body Hamiltonian that describes a lattice of
these atoms is \beqn H =  \sum_i \Omega_i \sigma^x_i + H_0, \ \
H_0 = - \sum_i \delta_i \hn_i + \sum_{ij} V_{ij} \hn_i \hn_j,
\label{H} \eeqn where $V_{ij} = C_{ij}/|i - j|^6$ and $i,j$ label
the lattice sites.

\begin{figure}
\begin{center}
\includegraphics[width=0.45\textwidth]{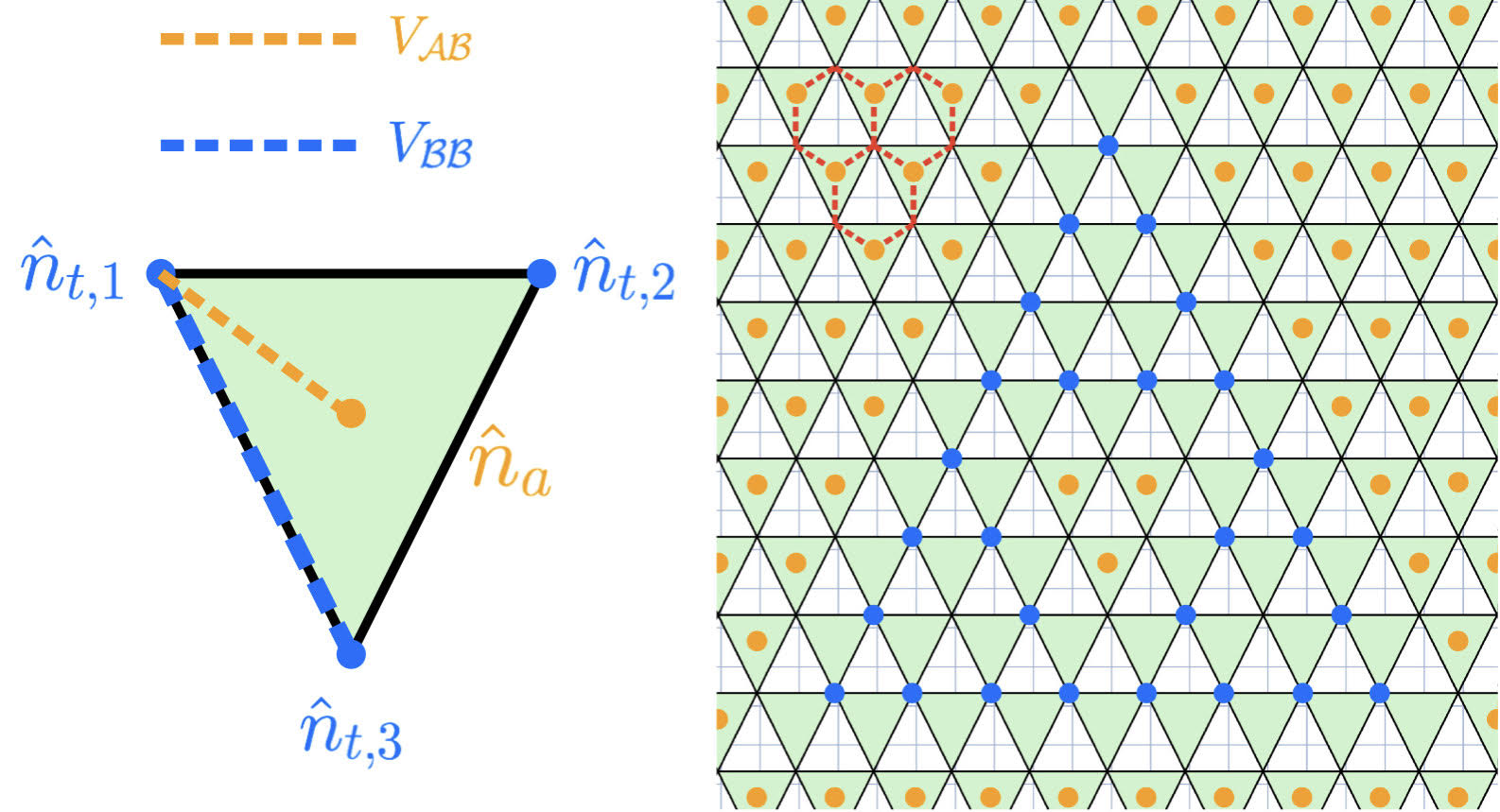}
\caption{We trap atoms on both the vertices and the center of each
downward triangle of a triangular lattice, which together form a
honeycomb lattice (three hexagons of the honeycomb lattice are
shown in the figure). The atoms on the vertices and the triangle
centers are the ``target" and ``auxiliary atoms" respectively. The
Hamiltonian Eq.~\ref{H0} reduces to $V_{\cA\cB} = 4V + v$ between
a auxiliary atom and its neighboring target atoms, and interaction
$V_{\cB\cB} = 2V$ between two neighboring target atoms. The ground
state of Eq.~\ref{H0} demands that, on each downward facing
triangle, either $\hn_a = 1$ (labelled orange) and all $\hn_t =
0$; or $\hn_a = 0$ and two out of three $\hn_t = 1$ (labelled
blue).} \label{Construction}
\end{center}
\end{figure}

We will start with the small Rabi frequency (relative to the
detuning) limit of this model so that we may first ignore the
$\sigma^x$ terms and focus on the classical part of the
Hamiltonian $H_0$. To realize the classical Sierpinski triangle
model of Eq.~\ref{st}, we need to select the parameters in $H_0$
which yield low energy states that can be mapped to those of the
Sierpinski triangle model. We consider the honeycomb lattice with
two sublattices $\cA$ and $\cB$. We trap an ``auxiliary'' atom at
each site in $\cA$ and a ``target'' atom at each site in $\cB$. An
equivalent picture is that we take the triangular lattice and
decorate each vertex with a target atom and the center of each
downward facing triangle with an auxiliary atom
(Fig.~\ref{Construction}). We aim to reproduce the states of
Eq.~\ref{st} $\textbf{only on}$ the $\cB$ sublattice with target
atoms. The auxiliary atoms enlarge the Hilbert space and hence the
states of model Eq.~\ref{st} with multi-spin interactions can be
reproduced through two-body interactions only in the low energy
subspace of the atomic system.

We assign different principal quantum numbers $n_{\cA}$ and
$n_{\cB}$ for the Rydberg states of the auxiliary atoms located on
sublattice $\mathcal{A}$ and target atoms on sublattice
$\mathcal{B}$. Crucially, with a proper choice of $n_{\cA}$,
$n_{\cB}$, as well as the detuning, the Hamiltonian $H_0$ can be
organized as $H_0 = \sum_{a \in \mathcal{A}} H_{0, a}$: \beqn
H_{0,a} &=& V \left(2 \hn_a + \hn_{t,1} + \hn_{t,2} + \hn_{t,3} -
2 \right)^2 + \sum_{i = 1}^3 v \hn_a \hn_{t,i} \cr\cr &\sim&
V\left(\sum_{i = 1}^3 4 \hn_a \hn_{t,i} + \sum_{i < j} 2
\hn_{t,i}\hn_{t,j} - 4 \hn_a - \sum_{i=1}^3 3 \hn_{t,i} \right)
\cr\cr &+& \sum_{i = 1}^3 v \hn_a \hn_{t,i} \cdots. \label{H0}
\eeqn The sum $H_0 = \sum_{a \in \mathcal{A}} H_{0, a}$ is over
all sublattice sites $\mathcal{A}$, and each term in the sum
involves an auxiliary atom ($\hn_a = 0,1$), as well as its three
neighboring target atoms ($\hn_{t,i} = 0,1$) which form a
downwards facing triangle on the honeycomb lattice. The second
line of Eq.~\ref{H0} uses the fact that $\hn_a^2 = \hn_a$,
$\hn_t^2 = \hn_t$ for $\hn = 0, 1$. $H_0$ contains a two-body
repulsive interaction $V_{\cA\cB} = 4V + v$ between the auxiliary
atom and its neighboring target atoms, as well as a repulsive
interaction $V_{\cB\cB} = 2V$ between two nearest neighbor target
atoms (Fig.~\ref{Construction}).
%(which are on second-neighbor sites of the original
%honeycomb lattice).

When $v > 0$, there are two classes of configurations of
$(\hn_{t,i}; \ \hn_a)$ on each downward triangle, both of which
are the ground states of $H_0$: \beqn && (1) \ \ \hn_a = 1, \
\hn_{t,i} = 0; \cr\cr \mathrm{or} && (2) \ \ \hn_a = 0, \
\text{two of } \hn_{t,i} = 1. \label{state1} \eeqn As a
comparison, the ground states of the classical Sierpinski triangle
model $H_{\mathrm{ST}}$ also have two types of ground states on
each downwards facing triangle: \beqn && (1) \ \ \sigma^z_1 =
\sigma^z_2 = \sigma^z_3 = + 1; \cr \cr \mathrm{or} && (2) \ \
\text{Two of } \sigma^z_i = -1. \label{state2}\eeqn Now the ground
states of $H_0$ can be one-to-one mapped to the ground states of
the classical Sierpinski triangle model, as long as we identify
$\sigma^z_i = 1 - 2\hn_{t,i}$. In fact in the appendix we will
show that, all the states of Eq.~\ref{st} (ground and excited
states) can be mapped one-to-one to the low energy subspace of
$H_0$ when $v > 0$.

%We will focus on the simplest case where the lasers are applied
%homogenously among all atoms so that the Rabi frequency and
%detuning are constant across the lattice.

The relation between $V_{\mathcal{A}\mathcal{B}}$ and
$V_{\mathcal{B}\mathcal{B}}$ can be tuned by choosing the
principal quantum numbers $n_\cA$ and $n_\cB$ properly. For
example, for potassium atoms, if we choose $n_\mathcal{A} = 76$
and $n_\mathcal{B} = 113$, then using the techniques in
Ref.~\cite{Sibalic_2017} and the fact that the interatomic
distances are related by $r_{\mathcal{B}\mathcal{B}} = \sqrt{3}
r_{\mathcal{A}\mathcal{B}}$, we found that both interactions are
repulsive and satisfy $V_{\cA\cB}/V_{\cB\cB} \sim 2.628$ ($v \sim
1.26 V$). Note that different choices are possible; our results
apply more broadly than just to this specific choice of atom and
principal quantum numbers.

In the real system, there are perturbations to Eq.~\ref{H0}. These
include other terms induced by the Van der Waals interaction, for
example the repulsion between Rydberg states on two neighboring
auxiliary atoms, whose strength $V_{\mathcal{A}\mathcal{A}}$
compared with $V_{\mathcal{B}\mathcal{B}}$ is
$V_{\mathcal{A}\mathcal{A}}/V_{\mathcal{B}\mathcal{B}} \sim 0.011$
using the example principal quantum numbers we chose above. The
repulsion $V_{\cB\cB}'$ between the target atoms on two second
neighbor $\mathcal{B}$ sites is also much weaker than $V_{\cB\cB}$
due to the rapid decay of the Van der Waals interaction with
distance. Another notable perturbation is the interaction
$V'_{\mathcal{A}\mathcal{B}}$ between an auxiliary atom and its
next-neighbor target atom. Compared with $V_{\cB\cB}$, the two
perturbations $V_{\cB\cB}'$ and $V'_{\mathcal{A}\mathcal{B}}$ are
\beqn \frac{V'_{\cB\cB}}{V_{\cB\cB}} = \frac{1}{(\sqrt{3})^6} \sim
0.037, \ \
\frac{V'_{\mathcal{A}\mathcal{B}}}{V_{\mathcal{B}\mathcal{B}}} =
\frac{1}{2^6}
\frac{V_{\mathcal{A}\mathcal{B}}}{V_{\mathcal{B}\mathcal{B}}} \sim
0.041. \eeqn

These perturbations shift the energy of the excited state of the
Sierpinski triangle shape. Let us consider an excited state with
flipped ``spins" ($\hn_t = 1$) on a Sierpinski triangle with side
length $L = 2^\ell$. In the ideal model of Eq.~\ref{st}, the
energy of this excited state does not depend on $L$ or $\ell$: all
the energy cost arises from the corners of the Sierpinski triangle
and the energy compared with the ground state is $E_\text{ex} = E
- E_{g} = 6K = 3 V = 3/2 \times V_{\cB\cB}$. However, in the real
system the leading order perturbations $V_{\mcal{A} \mcal{B}}',
V_{\mcal{A} \mcal{A}}$ and $V_{\mcal{B} \mcal{B}}'$ cause the
energy cost of creating a Sierpinski triangle to scale with its
size. In particular, the energy of the excitation relative to the
ground state with uniform $\hn_a = 1$ and $\hn_t = 0$ is estimated
to be (details are given in the appendix)
\begin{equation}
E_{\text{ex}}^\text{Ry} = (3 /2 - 0.1 + 0.47 \times 3^{\ell-3})
V_{\mcal{B} \mcal{B}} \label{ex}
\end{equation}
for Sierpinski triangles of $\ell \geq 2$. Since the actual energy
cost of creating a Sierpinski triangle increases with the size of
the triangle, the perturbations can no longer be ignored for large
enough Sierpinski triangles. Finite-size fractal excitations,
however, will still be observable.

{\it --- Experimental Proposals}

\begin{figure}
\begin{center}
\includegraphics[width=0.27\textwidth]{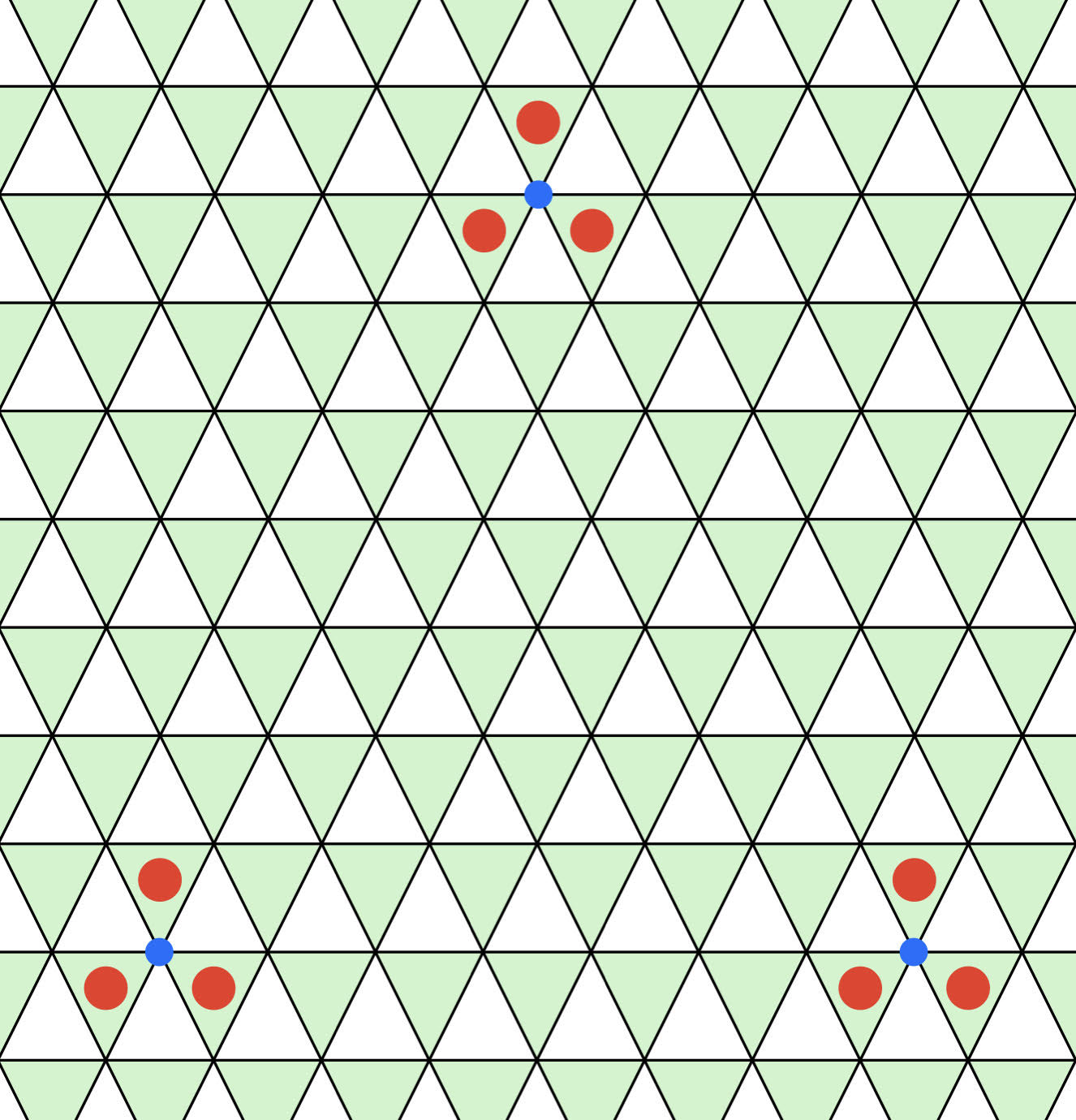}
\caption{With Hamiltonian Eq.~\ref{H0}, if three target atoms at
the corners of a triangle with side length $L = 8$ are excited
from the ground state to the Rydberg state, it costs energy $3V +
3v$ at each corner. If we apply an extra detuning $-
\bar{\delta}_{\mathrm{C}} \hn_t$ on the three corners of the
triangle, for sufficiently large $\bar{\delta}_{\mathrm{C}}$ the
ground state of the system is given by the fractal configuration
in Fig.~\ref{ST_model}.} \label{Corners}
\end{center}
\end{figure}

Next, we outline a procedure to enable experimental observation of
a spontaneously generated fractal-shaped excitation. This can be
done by adiabatically evolving a prepared ground state of
Eq.~\ref{H0} to the Sierpinski triangle excitation, which is the
ground-state of a new Hamiltonian achieved by slowly and carefully
varying the detuning and Rabi frequency. Note that this procedure
requires a level of local control beyond that present in most
current Rydberg experimental platforms, but the elements of which
have been demonstrated experimentally~\cite{PhysRevA.90.023415}.
We start with the Hamiltonian of Eq.~\ref{H0} with a small and
finite Rabi frequency. To ensure a unique ground state (to enable
an adiabatic evolution), we first deform Eq.~\ref{H0} with a small
extra detuning $\bar{\delta} \hn_t$ on all target atoms {\bf
outside of} a triangle with side $L = 2^\ell$. We then prepare an
initial state with all target atoms $\hn_t = 0$ and $\hn_a = 1$,
which is equivalent to $\sigma^z = 1$ uniformly in Eq.~\ref{st}
and represents the unique ground state of the Hamiltonian prepared
above. We slowly deform the Hamiltonian with time-dependent Rabi
frequency $\Omega(t)$ and detuning $\delta_i(t)$ within the
triangle, reaching the final Hamiltonian with extra detuning $-
\bar{\delta}_{\mathrm{C}} \hn_t$ localized to three target atoms
at the corners of the triangle (Fig.~\ref{Corners}). With
sufficiently large $\bar{\delta}_{\mathrm{C}}$, the unique ground
state of the final Hamiltonian contains a Sierpinski triangle
arrangement of all the atoms inside the triangle, as shown in
Fig.~\ref{ST_model} and Fig.~\ref{Construction}, despite the fact
that the extra detuning $- \bar{\delta}_{\mathrm{C}} \hn_t$ is
only applied locally at the corners rather than throughout the
entire interior. Based on our estimate of energy in Eq.~\ref{ex}
which includes further leading neighbor interactions arising from
the Van der Waals interaction, this phenomenon can hold up to
Sierpinski triangles with $\ell = 4$ (side length $L = 16$,
containing 81 atoms), as a single Sierpinski triangle
configuration still has lower energy than fragmented
configurations.

We have specified the initial and final Hamiltonian for the
desired evolution; an adiabatic path of the detuning and Rabi
frequency for observing Ising-like crystallization of Rydberg
atoms has already been
demonstrated~\cite{adiabatic,browaeys2daf,lukinphases}. Leveraging
the aforementioned power of local control, we could also create
arbitrary configurations of the Rydberg array, including the
Sierpinski triangle. Spectroscopic measurement of the energy of
the Sierpinski configuration compared with that of fragmented
configurations could then confirm that it represents a low-lying
excited state. This technique would no longer rely on adiabatic
evolution.

This experimental platform also gives us the potential to probe
quantum phase transitions by controlling the Rabi frequency. When
the Rabi frequency $\Omega$ is increased uniformly on all the
target atoms, around $\Omega \sim V$ there is expected to be a
quantum phase transition similar to the one recently studied
numerically~\cite{juan1,frankyizhi}. In the large $\Omega$ phase,
one is not supposed to observe the fractal configuration in the
proposed experiment above, i.e. the configuration of
Fig.~\ref{Corners} will not evolve to the one in
Fig.~\ref{ST_model} and Fig.~\ref{Construction}. As we pointed out
before, the nature of this transition is far from being
understood. Hence the realization of the quantum phase transition
in experimental systems with high tunability is crucial to
understand these exotic quantum phase transitions, and to test
possible theoretical paradigms developed in the future.

The fractal structure of the fractal orders also manifests at the
level of the correlation functions. In both the classical and
quantum Sierpinski triangle models, the three-point correlation
function $\mathcal{C}_3 = \langle \sigma^z_{j}
\sigma^z_{j+L\hat{x}} \sigma^z_{j+L(\hat{x} - \hat{y})} \rangle$
is a characteristic quantity which plays the role of the
correlation function of ordinary quantum many-body systems. The
three-point correlation decays hyper-exponentially with the
Hausdorff dimension at finite temperature~\cite{yoshida,NM} and
its scaling at the quantum critical point $h = K$ was computed in
Ref.~\cite{frankyizhi}. In the experimental realization of
Eq.~\ref{H0}, the three-point correlation function $\mcal{C}_3$
can be reconstructed by averaging over multiple single-site
resolution snapshots of the configuration of the target atoms
taken in separate experimental realizations. Similar techniques
have been used previously in cold-atom experiments to reconstruct
quantities such as the spin correlation functions in Fermi-Hubbard
systems~\cite{Parsons1253,Endres200}. Experimental measurement of
$\mcal{C}_3$, whose scaling with distance diagnoses the fractal
physics of the model, will be crucial in understanding the nature
of this quantum phase transition.

{\it --- Discussion}

Previous proposals for realizing fracton related states mostly
focused on states that are analogous to type-I
fractons~\cite{fractonreal1,fractonreal2,fractonreal3,fractonreal4}.
For example, theoretical efforts on constructing fracton related
states were made based on localized Majorana zero modes for both
type-I and type-II states~\cite{fractonreal5,fractonreal6}.
Compared with previous proposals, the platform of Rydberg atoms
discussed in the current work is highly tunable with precision at
the level of a single atom. Another advantage of the platform of
Rydberg atoms, and in general cold atom systems, is fast
manipulation of parameters in the Hamiltonian, which can either
periodically drive the system, or cause a quantum
quench~\cite{polkovnikov-noneqRMP,atomfloquet1,atomfloquet2,prethermPRX,atomquench,Bluvstein1355,KZrydberg,berniendynamics}.
Many exotic features are expected in the quantum dynamics of
fracton related models due to the restricted motion of the fracton
excitations~\cite{fractondyna1,fractondyna2,fractondyna3,fractondyna4}.
By quickly tuning parameters such as the Rabi frequency in
Eq.~\ref{H}, one can compare the quantum dynamics of the fractal
order simulated with Rydberg atoms with future analytical and
numerical analysis. Furthermore, our construction of Eq.~\ref{H0}
used to reproduce multi-spin interactions in Rydberg Hamiltonians
with only two-body interactions but an enlarged Hilbert space, can
be extended to fracton related models richer than the Sierpinski
triangle model. One example of such extension is given in the
appendix.

The authors thank Chao-Ming Jian and Hannes Bernien for very
helpful discussions. C.X. is supported by NSF Grant No.
DMR-1920434, and the Simons Investigator program. D.W.
acknowledges support from the Army Research Office (MURI
W911NF1710323), and the University of California Multicampus
Research Programs and Initiatives (MRP19-601445). We gratefully
acknowledge support via the UC Santa Barbara NSF Quantum Foundry
funded via the Q-AMASE-i program under award DMR-1906325. This
material is based upon work supported by the U.S. Department of
Energy, Office of Science, National Quantum Information Science
Research Centers, Quantum Science Center.

\bibliography{Rydberg}

\appendix

\section{Estimate of energy of Sierpinski triangle excitation in real system}

With the inclusion of the perturbations of $V_\mcal{AA},
V_\mcal{BB}'$ and $V_\mcal{AB}'$, we can calculate the relative
cost of creating a Sierpinski triangle excited state by exciting
target atoms and de-exciting auxiliary atoms in the perturbed
ground state with $\hn_a = 1$ and $\hn_t = 0$ uniformly. For a
Sierpinski triangle of length $L = 2^\ell$, the energy cost
$E_\text{ex}^\text{Ry}$ is \beqn E_\text{ex}^\text{Ry} = 3 V +
N'_{\mcal{BB}} V_\mcal{BB}' +  N'_\mcal{AB} V_\mcal{AB}' -
N_\mcal{AA} V_\mcal{AA}, \eeqn where $N'_\mcal{BB}$,
$N'_\mcal{AB}$ and $N_\mcal{AA}$ are the number of interactions
$V_\mcal{BB}'$, $V_\mcal{AB}'$ we have turned on, and the number
of interactions $V_\mcal{AA}$ we have turned off, respectively, in
order to create the Sierpinski triangle.

\begin{figure}
\begin{center}
\includegraphics[width=0.3\textwidth]{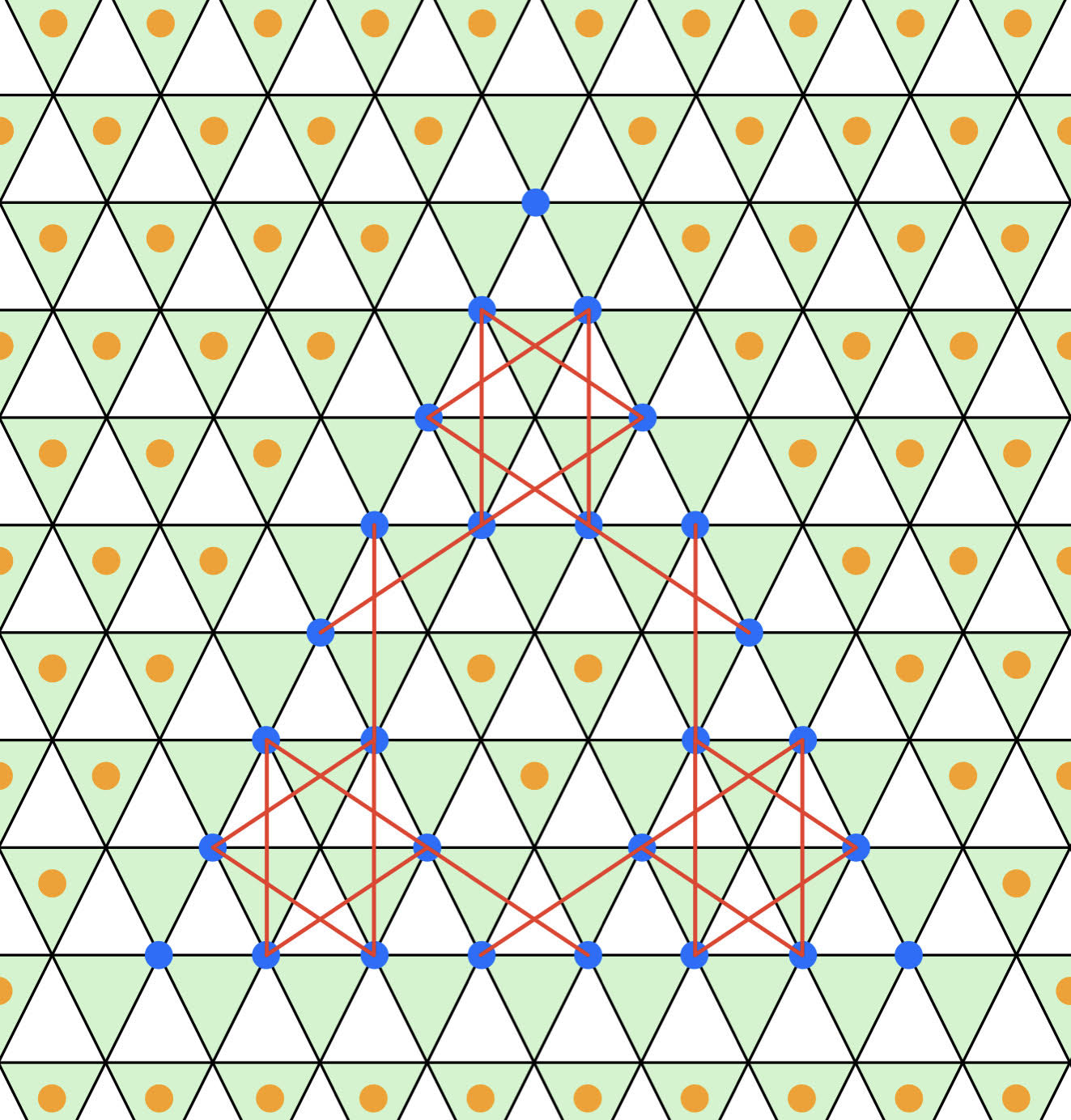}
\caption{The next neighbor Van der Waals interaction between
target atoms.}\label{VBB'}
\end{center}
\end{figure}

\begin{figure}
\begin{center}
\includegraphics[width=0.3\textwidth]{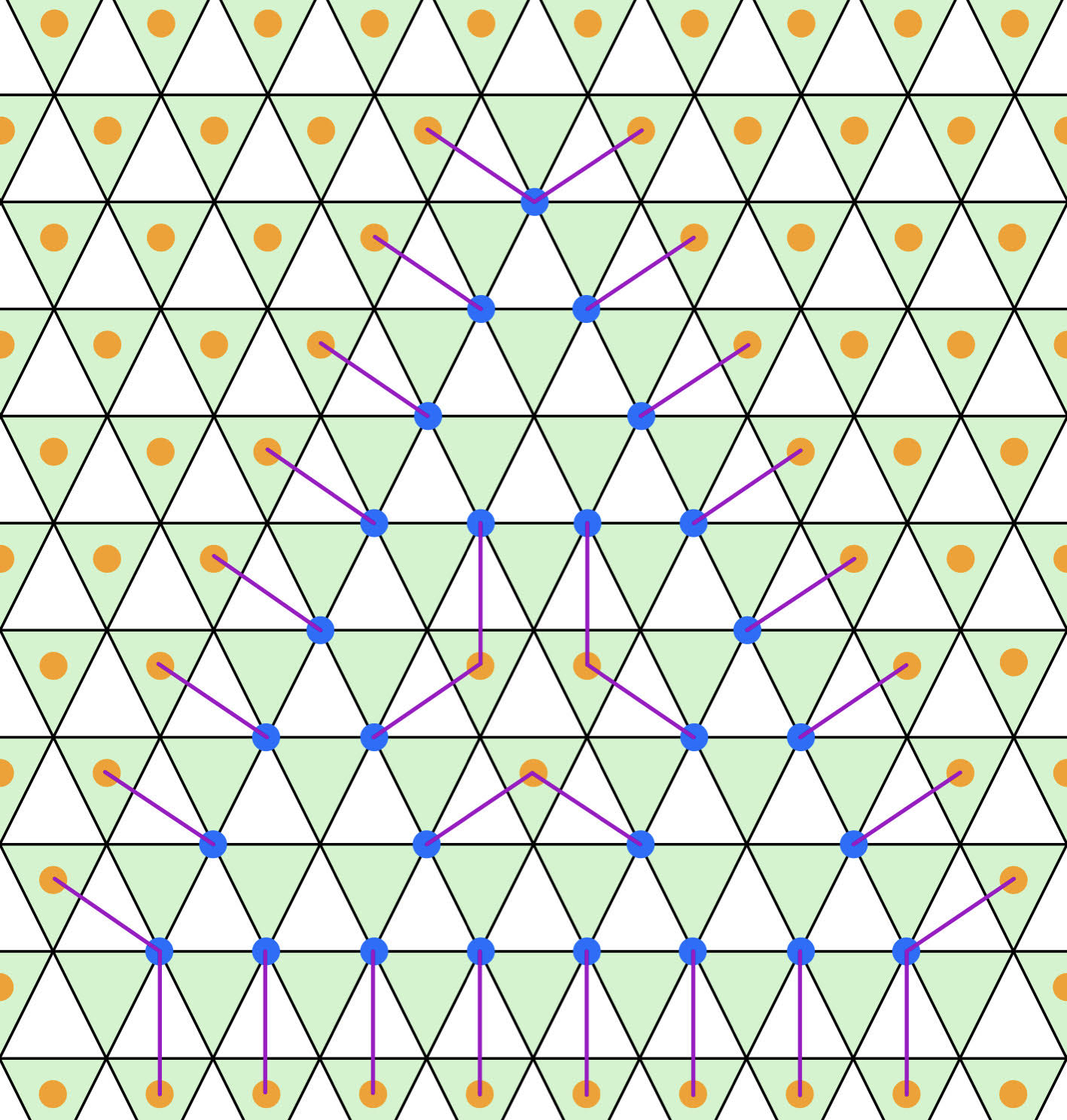}
\caption{The next neighbor interaction between auxiliary atoms and
target atoms.}\label{VAB'}
\end{center}
\end{figure}

\begin{figure}
\begin{center}
\includegraphics[width=0.3\textwidth]{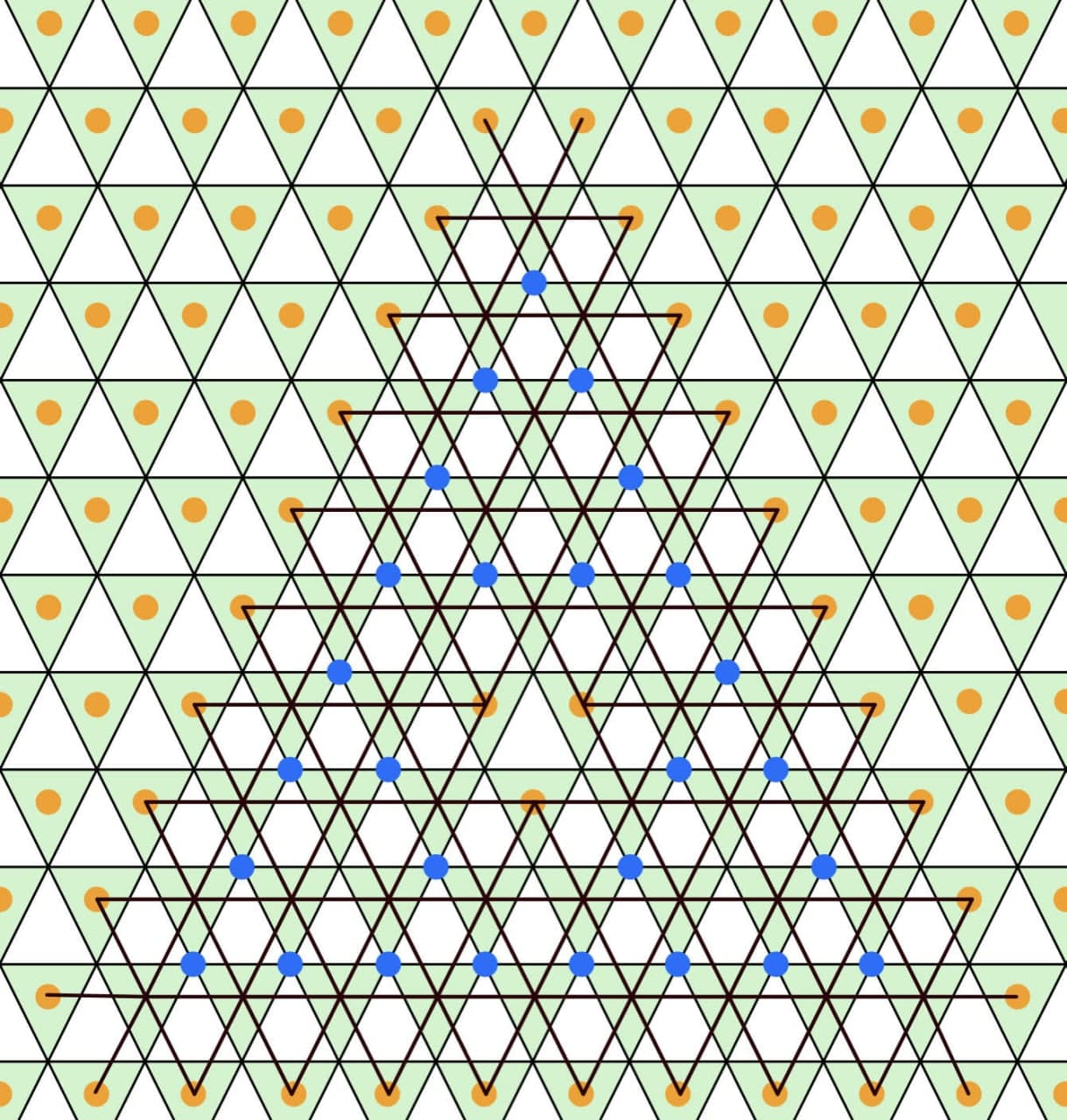}
\caption{The nearest neighbor interaction between auxiliary atoms
that are ``lost" while creating the Sierpinski triangle
excitation.}\label{VAA}
\end{center}
\end{figure}

Here $N_\mcal{BB}'$ as a function of $\ell$ is \beqn
N'_\mcal{BB}(\ell) = (3^\ell - 3) \text{ for }\ell \geq 2. \eeqn
This result can be seen from the following: each Sierpinski
triangle of length $2^{\ell - 1}$ has a corresponding number of
$V_{\mcal{BB}}'$ interactions $N'_{\mcal{BB}}(\ell - 1)$. The
Sierpinski triangle of length $2^\ell$ is constructed by gluing
three copies of of the smaller Sierpinski triangle of length
$2^{\ell - 1}$ together at the corners so that it has
$3N'_{\mcal{BB}}(\ell - 1)$ many interactions, as well as the
added interactions between target atoms belonging to different
smaller Sierpinski triangles (Fig.~\ref{VBB'}). Including these
added interactions, of which there are $6$, we recover a
recurrence relation for $N'_{\mcal{BB}}(\ell)$: \begin{equation*}
N'_{\mcal{BB}}(\ell) = 3N'_{\mcal{BB}}(\ell-1) + 6 \end{equation*}
\noindent with the number of interactions for the $\ell = 2$
Sierpinski triangle as the initial condition, $N'_{\mcal{BB}}(2) =
6$. This recurrence relation can be solved for
\begin{equation*} N'_{\mcal{BB}}(\ell) = (3^\ell-3) \text{ for } \ell \geq 2,
\end{equation*}
\noindent which is the stated result.

We can compute the number of $V_\mcal{AB}'$ interactions that we
have turned on in a similar manner. The total number of these
interactions is
\begin{equation*}
N'_\mcal{AB}(\ell) = (3^\ell + 3) \text{ for } \ell \geq 2.
\end{equation*}
\noindent Since these are interactions between auxiliary atoms and
target atoms, these occur between the target atoms on the edges of
the Sierpinski triangle and surrounding auxiliary atoms, as well
as between target atoms and the auxiliary atoms in the interior of
the Sierpinski triangle. For a Sierpinski triangle of length $L =
2^\ell$, the total number of interactions coming from the target
atoms at the edges is $3L = 3 \times 2^\ell$. Counting the
interactions coming from the interior of the Sierpinski triangle
$N_{\mcal{AB}}^{\text{int}}(\ell)$ is more complicated as it
depends on the length of the previous iteration of the Sierpinski
triangle. These can be counted as the contribution from the
largest missing triangle $3 \times(2^{\ell - 1} - 2)$, plus three
times the number of interactions from the last iteration
$3N_{\mcal{AB}}^{\text{int}}(\ell -1)$. As a result, we are yet
again left with a recurrence relation
\begin{equation*}
N_{\mcal{AB}}^{\text{int}}  (\ell) = 3N_{\mcal{AB}}^{\text{int}}
(\ell-1) + 3 \times (2^{\ell-1} - 2)
\end{equation*}
\noindent with initial condition $N_{\mcal{AB}}^{\text{int}}(2) =
0$. This can be solved for
\begin{align*}
N_{\mcal{AB}}^{\text{int}}  = 3^\ell - 3 \times 2^\ell + 3.
\end{align*}
\noindent Therefore the total number of $V_{\mcal{AB}}'$
interactions we have turned on is
\begin{align*}
N_{\mcal{AB}} &= N_{\mcal{AB}}^{\text{int}}  + 3 \times 2^\ell,\\
&= (3^\ell + 3) \text{ for } \ell \geq 2.
\end{align*}

The calculation of the number of $V_\mcal{AA}$ interactions lost
when creating the Sierpinski triangle excitation is nearly
identical to the previous calculation. The total number of lost
interactions can be calculated as \beqn N_\mcal{AA} = \frac{1}{2}
( 11 \times 3^\ell +21)\text{ for } \ell \geq 2. \eeqn

There are three sources of the loss of $V_\mcal{AA}$ interactions:
1) the loss of interaction between two de-excited auxiliary atoms
in downwards triangles in the Sierpinski triangle, 2) the loss of
interaction between a de-excited auxiliary atom on the edges and
corners of the Sierpinski triangle and an excited auxiliary atom
surrounding the Sierpinski triangle and 3) the loss of interaction
between a de-excited auxiliary atom in the Sierpinski triangle and
an excited auxiliary atom in the interior of the Sierpinski
triangle.

The contribution to the total number of lost $V_\mcal{AA}$
interactions from the first two cases 1) and 2) can be found to be
$\frac{1}{2}(7 \times 3^{\ell}-3) + 6(2^\ell +1)$. There are an
additional $12$ lost interactions that come from de-excited
auxiliary atoms at the corners of the Sierpinski triangle and do
not scale with $\ell$. For the $\ell = 2$ Sierpinski triangle
there are $N_{\mcal{AA}}^{1)}(2) = 30$ lost interactions not
including the ones from the edges and corners. Gluing these
triangles to generate a $\ell = 3$ Sierpinski triangle gives us
$3$ times as many interactions as well as an additional $3$ lost
interactions between auxiliary atoms belonging to different
smaller Sierpinski triangles. Hence we are left with a recursion
relation
\begin{equation*}
N^{1)}_{\mcal{AA}}(\ell) = 3N^{1)}_{\mcal{AA}}(\ell-1) + 3.
\end{equation*}
Using $N_{\mcal{AA}}^{1)}(2) = 30$ as the initial condition, we
have a solution of
\begin{equation*}
N_{\mcal{AA}}^{1)}(\ell) = \frac{1}{2}(7 \times 3^\ell - 3) \text{
for } \ell \geq 2.
\end{equation*}
$N_{\mcal{AA}}^{2)}$ is straightforward: $N_{\mcal{AA}}^{2)} =
6(2^\ell - 1)+12 = 6(2^\ell+1)$ including the interactions at the
corners. Hence, restoring the additional $12$ interactions,
$N_{\mcal{AA}}^{1)}+N_{\mcal{AA}}^{2)} = \frac{1}{2}(7 \times
3^{\ell}-3) + 6(2^\ell +1)$.

We recover another recurrence relation for the number of lost
interactions in the interior for case 3):
\begin{equation*}
N_{\mcal{AA}}^{3)}(\ell) = 3 N_{\mcal{AA}}^{3)}(\ell-1) + 6 \times
(2^{\ell - 1} - 2)
\end{equation*}
with initial condition $N_{\mcal{AA}}^{3)}(2) = 0$. This can be
solved for
\begin{equation*}
N_{\mcal{AA}}^{3)}(\ell) = 2 \times 3^\ell - 3 \times 2^{\ell + 1}
+ 6.
\end{equation*}
The total number of lost $V_\mcal{AA}$ interactions is then
\begin{align*} N_{\mcal{AA}} &= N_{\mcal{AA}}^{3)}(\ell) +
\frac{1}{2}(7 \times 3^{\ell}-3) + 6(2^\ell +1), \\ &= \frac{1}{2}
( 11 \times 3^\ell +21)\text{ for } \ell \geq 2.
\end{align*}

Combining all the perturbations evaluated above,
$E_\text{ex}^\text{Ry}$ is given by \beqn E_\text{ex}^\text{Ry}
&=3 V +  N'_{\mcal{BB}} V_\mcal{BB}' +  N'_\mcal{AB} V_\mcal{AB}'
-  N_\mcal{AA} V_\mcal{AA},\cr\cr &\approx (3 /2 - 0.1 + 0.47
\times 3^{\ell-3}) V_\mcal{BB}. \eeqn

\section{Mapping between states}

On a unit downward facing triangle, the mapping between states in
Eq.~\ref{st} and the states in Eq.~\ref{H0} is: \beqn && \sigma^z
= (+1, +1, +1), \ \ \mathrm{Energy} = - K \cr\cr \rightarrow &&
(\hn_{t,i}; \hn_a) = (0, 0, 0; 1), \ \ \mathrm{Energy} = 0; \cr\cr
&& \sigma^z = (-1, -1, +1), \ \ \mathrm{Energy} = - K, \cr\cr
\rightarrow && (\hn_{t,i}; \hn_a) = (1, 1, 0; 0), \ \
\mathrm{Energy} = 0; \cr\cr && \sigma^z = (-1, +1, +1), \ \
\mathrm{Energy} = +  K, \cr\cr \rightarrow && (\hn_{t,i}; \hn_a) =
(1, 0, 0; 0), \ \ \mathrm{Energy} = V; \cr\cr && \sigma^z = (-1,
-1, -1), \ \ \mathrm{Energy} = + K \cr\cr \rightarrow &&
(\hn_{t,i}; \hn_a) = (1, 1, 1; 0), \ \ \mathrm{Energy} = V. \eeqn
The Hilbert space of the atoms is bigger than the one of the
spins; hence there are some extra states with even higher energy.
The $v \hn_a \hn_{t,i}$ term in $H_0$ does not affect the energy
of the states list above, but they will separate the other states
in the atomic Hilbert space with extra energy cost: \beqn &&
(\hn_{t,i}; \hn_a) = (1, 0, 0; 1), \ \ \mathrm{Energy} = V + v;
\cr\cr && (\hn_{t,i}; \ \hn_a) = (1, 1, 0; 1), \ \ \mathrm{Energy}
= 4V + 2 v; \cr\cr && (\hn_{t,i}; \ \hn_a) = (1, 1, 1; 1), \ \
\mathrm{Energy} = 9V + 3 v ; \cr\cr && (\hn_{t,i}; \ \hn_a) = (0,
0, 0; 0), \ \ \mathrm{Energy} = 4V. \eeqn Hence with positive $v$,
the low energy states of the Rydberg atom systems can be exactly
mapped to the states of the Ising spins in Eq.~\ref{H0}, with $V =
2K$.

\section{Fractal symmetry in the quantum Sierpinski triangle model}

\begin{figure}
\includegraphics[width=0.48\textwidth]{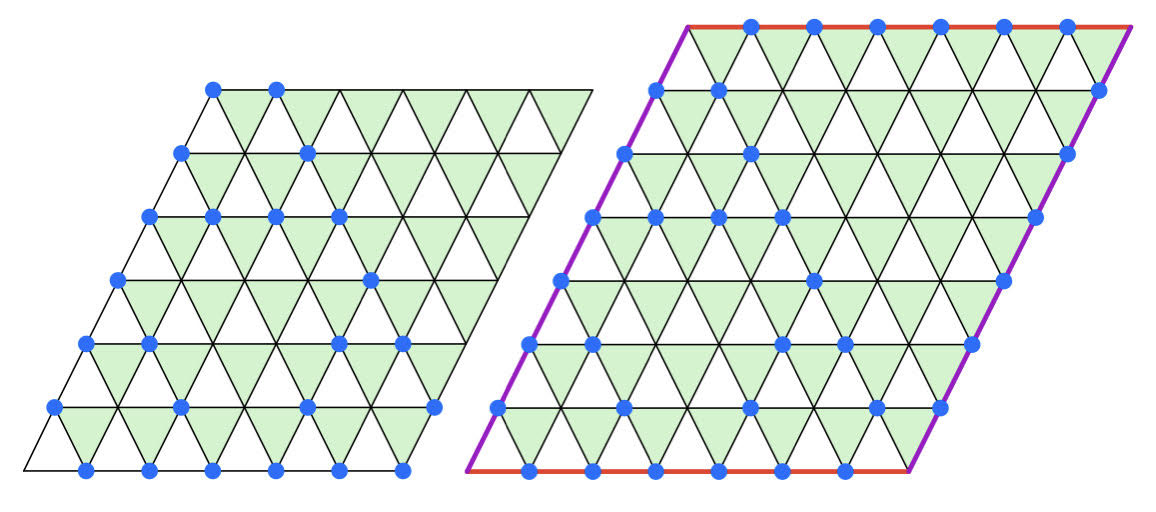}
\caption{(Left): when the system is defined on a triangular
lattice with periodic boundary conditions, and $L_x = L_y = 2^k -
1$ ($k = 3$ in the figure), there exist symmetry generation
operators $\hat{G}(\mathcal{L}) = \prod_{j \in \mathcal{L}}
\sigma^x_j$ where $\mathcal{L}$ are all the sites labeled in blue.
(Right): we duplicate the bottom row on the top, and the left
column on the right of the lattice to demonstrate the periodic
boundary conditions. }\label{periodic}
\end{figure}

For a quantum system to have a symmetry, the Hamiltonian of the
system needs to commute with a symmetry generation operator
$\hat{G}$. For example, the standard quantum Ising model $H =
\sum_{\langle i,j \rangle} - K \sigma^z_i \sigma^z_j - \sum_j h
\sigma^x_j$ has a $Z_2$ spin symmetry that is generated by the
operator $\hat{G} = \prod_j \sigma^x_j$ which commutes with the
entire Hamiltonian. Although the Hamiltonian has $Z_2$ spin
symmetry, in the Ising ordered phase with $K \gg h$, (at least in
a certain basis) the ground state wave function of the system in
the thermodynamics limit is not invariant under the operation of
$\hat{G}$. The ground states are degenerate in the ordered phase,
and $\hat{G}$ would take one ground state to another. In this
sense, the $Z_2$ symmetry is spontaneously broken in the Ising
ordered phase. On the other hand, in the Ising disordered phase $h
\gg K$ the $Z_2$ symmetry is preserved,i.e. the ground state wave
function is unique and invariant under operation of $\hat{G}$.

In the quantum Sierpinski triangle model Eq.~\ref{qst}, let us
consider periodic boundary conditions for convenience. The
Hamiltonian does not have exact subsystem symmetry for most system
sizes; however, when the system is a parallelogram with side
length $L_x = L_y = 2^k - 1$ and $k$ being a positive integer,
there exists a series of symmetry generation operators
$\hat{G}(\mcal{L})$ which are a product of $\sigma^x_j$ on a
subset of the lattice that belong to a Sierpinski triangle:
\begin{equation}
\hat{G}(\mcal{L}) = \prod_{j \in \mcal{L}} \sigma_j^x.
\end{equation}
The shape of $\mcal{L}$ is illustrated in Fig.~\ref{periodic}.
Notice that due to the subtlety of the periodic boundary
conditions, $\hat{G}(\mcal{L})$ commutes with the entire
Hamiltonian when and only when the system size $L_x = L_y = 2^k -
1$. And when $K > h$, in the thermodynamics limit $(k \rightarrow
\infty)$ the system has many degenerate ground states which are
not invariant under the operation of $\hat{G}(\mcal{L})$, hence
the phase $K > h$ spontaneously breaks the fractal symmetry
generated by $\hat{G}(\mcal{L})$. While in the phase $h > K$ the
ground state is unique and invariant under operation of
$\hat{G}(\mathcal{L})$.

\section{The Sierpinski tetrahedron model}

\begin{figure}
\includegraphics[width=0.48\textwidth]{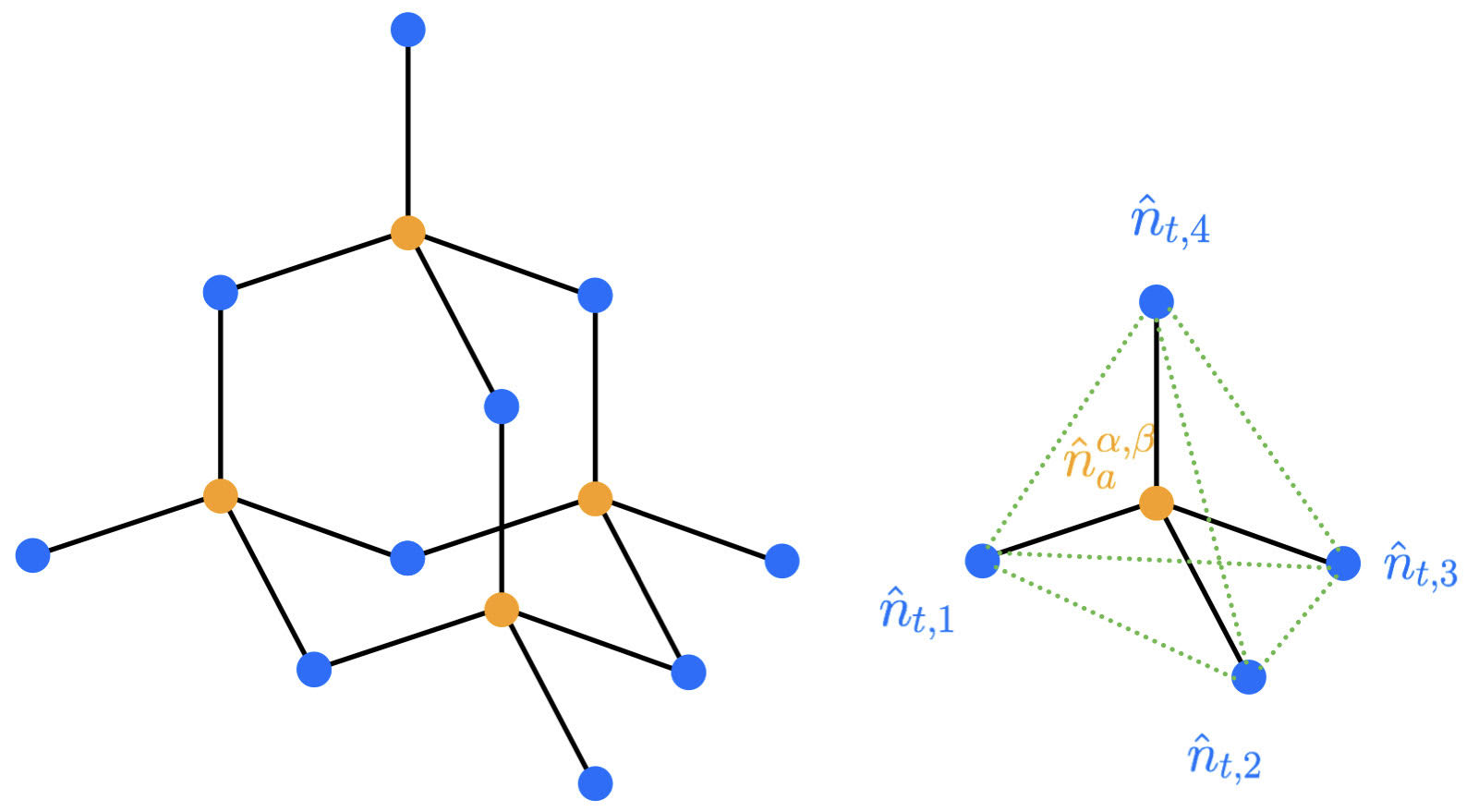}
\caption{The diamond lattice constructed out of tetrahedra with
target atoms $\hn_{t,1}, \dots, \hn_{t,4}$ decorating the vertices
and auxiliary atoms supporting two different quantum numbers
$\hn_a^{\alpha, \beta}$ decorating the centers.}\label{tetrafig}
\end{figure}

One of the main obstacles in constructing many exotic states of
matter considered theoretically is the multi-spin interaction in
the Hamiltonian. Our platform constructed with Rydberg atoms can
simulate much of the low energy physics of the Sierpinski triangle
model, which originally involves three-body interactions. In fact,
our construction can be expanded to other fractal models with more
general multi-spin interactions as well. An example of such a
model that we might hope to realize is the natural extension of
the Sierpinski triangle model to three-dimensions: the Sierpinski
tetrahedron model~\cite{juan2,stetra}. To construct this model, we
first glue tetrahedra with effective spin-$1/2$ degrees of freedom
$(\sigma^z = \pm 1)$ at the vertices together at the corners to
form a lattice. As each tetrahedron has four vertices, one may sum
over all four-body interactions on each tetrahedron,
\begin{equation}
H_{\text{tetra}} = \sum_\text{tetra} -K \sigma_1^z \sigma_2^z
\sigma_3^z \sigma_4^z .\label{tetra}
\end{equation}
Many of the technical features of this model are similar to those
of the Sierpinski triangle model of Eq.~\ref{st}:

(1) The low energy excitations above the ground state are created
at the corners of a Sierpinski tetrahedron (a self-similar fractal
with Hausdorff dimension $d_H = 2$) of flipped spins and are
fractons.

(2) The quantum model with an added transverse field $\sum_j - h
\sigma^x_h$ is self-dual and there is a fractal symmetry that is
spontaneously broken in the phase $K > h$. A quantum phase
transition is expected at $h = K$, although the nature of the
transition needs further study.

Fractal-ordered phases such as the Sierpinski triangle and
Sierpinski tetrahedron models tend to have their fractal structure
manifest at the level of the correlation functions. In the
Sierpinski triangle model, the three-point correlation function
$\mathcal{C}_3 = \langle \sigma^z_{j} \sigma^z_{j+L\hat{x}}
\sigma^z_{j+L(\hat{x} - \hat{y})} \rangle$ is known to be
non-vanishing only when $L = 2^\ell$ and decays
hyper-exponentially with the Hausdorff dimension at finite
temperature~\cite{yoshida,NM}.

In the Sierpinski tetrahedron model, the four-point correlation
function $\mathcal{C}_4 = \langle \sigma^z_j \sigma^z_{j +
L\hat{x}} \sigma^z_{j+L\hat{y}} \sigma^z_{j + L \hat{z}} \rangle$
displays the same smoking-gun signatures of fractal order as
$\mcal{C}_3$ in the Sierpinski triangle model. This correlation
function may be computed in a similar fashion as one might compute
the two-point correlation function of the $1d$ Ising model,
$\langle \sigma^z_j \sigma^z_{j+L} \rangle$. Since there is a
one-to-one correspondence between tetrahedra and vertices, we
define new spin-1/2 variables at the center of each tetrahedron
$\tau^z_T = (\sigma^z_1 \sigma^z_2 \sigma^z_3 \sigma^z_4)$. The
Hamiltonian now describes a collection of statistically
independent spin variables $\tau_T$,
\begin{equation}
H_\text{tetra,d} = \sum_{T \in \text{tetra}} -K \tau_T.
\end{equation}
Since the $\tau$ variables are independent from one another, the
correlations of products of $\tau$'s factor nicely such that \beqn
\bigg  \langle \prod_j \tau_j \bigg \rangle &=& \prod_j \langle
\tau_j \rangle, \cr\cr &=& \prod_j \frac{\sum_{\{ \tau \}} \tau_j
e^{\beta K \sum_T \tau_T}}{\sum_{\{ \tau \}} e^{\beta K \sum_T
\tau_T}},\cr\cr &=& \prod_j \bigg ( \frac{e^{\beta K} - e^{- \beta
K}}{e^{\beta K} + e^{- \beta K}} \bigg ),\cr\cr &=&  \tanh(\beta
K)^{\# \text{ of } \tau \text{'s}}. \eeqn \noindent Taking
products of spins over a Sierpinski tetrahedron of length $L =
2^\ell$ for integer $\ell$ allows us to write the four-point
correlation as \beqn \mathcal{C}_4  &=& \bigg \langle \prod_{j \in
\text{ST}(L)} \tau_j \bigg \rangle , \cr\cr &=& \tanh(\beta K)^{\#
\text{ of tetrahedra}},\cr\cr &=& \tanh(\beta K)^{L^2}. \eeqn The
four point correlations for $L \neq 2^\ell$ must vanish.

The approach we have taken in realizing the Sierpinski triangle
model in Eq.~\ref{H0} can be extended to this Sierpinski
tetrahedron model with a four-body interaction. Starting from the
diamond lattice, we decorate each vertex and center of every
tetrahedron with target atoms and an auxiliary atom respectively
as shown in Fig.~\ref{tetrafig}. We couple the auxiliary atom to
two different Rydberg states using lasers with different
frequencies, and the two Rydberg states are labelled by their
principal quantum numbers $n^\alpha$ and $n^\beta$ ($n^\alpha >
n^\beta$ without loss of generality). Then we propose the
following Hamiltonian: \beqn H_0 \sim \sum_{a} V \bigg (4
\hn_a^\alpha + 2\hn_a^\beta + \sum_{v \in \text{tetra}}
\hat{n}_{t,v} - 4 \bigg )^2 + \dots \label{H02} \eeqn \noindent
where the ellipsis include all the longer-range couplings of
auxiliary and target atoms and the second sum is over all vertices
of the tetrahedron with $\hn_a^{\alpha,\beta}$ at its center.
Again, as we did for the Sierpinski triangle, we may select
principal quantum numbers $n^\alpha_a$, $n^\beta_a$, and $n_{t,i}$
such that the ratio of Van der Waals interaction strengths yields
our desired Hamiltonian Eq.~\ref{H02}. In Eq.~\ref{H02}, the
expansion will formally include terms such as $(\hn_a^\alpha)^2$,
$(\hn_a^\beta)^2$ and $\hn_{t,v}^2$, but these terms can be
absorbed by shifting the detuning as $\hn^2 = \hn$ for $\hn = 0,
1$. The expansion of Eq.~\ref{H02} also contains an interaction
between $\hn_a^\alpha$ and $\hn_a^\beta$ which seems unphysical.
However this terms does not change the energy of any physical
states because $\hn^\alpha$ and $\hn^\beta$ cannot be both nonzero
at the same site.

The ground states of Eq.~\ref{H02} are found when each term in the
sum vanishes. The possible configurations of $\hn$ on each
tetrahedra when this occurs are \beqn && (1) \ \ \hn_a^\alpha = 1,
\ \hn_a^\beta = 0, \ \hn_{t,i} = 0; \cr\cr  \mathrm{or} && (2) \ \
\hn_a^\alpha = 0, \ \hn_a^\beta = 1, \ \text{two of }\hn_{t,i} =
0; \cr\cr \mathrm{or} && (3) \ \ \hn_a^\alpha = 0;  \ \
\hn_a^\beta = 0; \ \ \hn_{t,i} = 1. \eeqn \noindent These ground
states of each tetrahedron of Eq.~\ref{H02} can be one-to-one
mapped to the ground states of Eq.~\ref{tetra}: (1) $\sigma^z_1 =
\sigma^z_2 = \sigma^z_3 = \sigma^z_4 = 1$, or (2) two of $
\sigma^z_i = -1$ or (3) $\sigma^z_1 = \sigma^z_2 = \sigma^z_3 =
\sigma^z_4 = -1$, under the correspondence $\sigma^z_i = 1 - 2
\hn_{t,i}$. Therefore our platform provides a natural extension to
the modeling of classical fracton models and multi-spin
interactions beyond the Sierpinski triangle model.

\end{document}